\documentclass[%
pra,
reprint,
superscriptaddress,
amsmath,amssymb,
aps,
]{revtex4-2}

\usepackage{times}
\usepackage{graphicx}
\usepackage{dcolumn}
\usepackage{bm}
\usepackage{amsmath,amssymb}
\usepackage{physics}
\usepackage[normalem]{ulem}
\usepackage{url,hyperref}
\usepackage{fixme}
\usepackage{extarrows}
\usepackage{xcolor}

 \fxsetup{status=draft}
 \fxsetup{inline=false, margin=true}
 \fxsetup{theme=color, author=GF}
\graphicspath{{./pics/}}
\DeclareGraphicsExtensions{.png,.jpg,.eps,.pdf}

\begin{document}
	
	\preprint{APS/123-QED}
	
	\title{Single-atom maser with engineered circuit for population inversion}
	
	\author{A.A. Sokolova}
	\email{Sokolova.aa@phystech.edu}
	\affiliation{Russian Quantum Center, Skolkovo village, Russia}
	\affiliation{Moscow Institute of Physics and Technology, Dolgoprundiy, Russia}
	\affiliation{National University of Science and Technology MISIS, Moscow, Russia}
	
	\author{G.P. Fedorov}
	\affiliation{Russian Quantum Center, Skolkovo village, Russia}
	\affiliation{Moscow Institute of Physics and Technology, Dolgoprundiy, Russia}
	\affiliation{National University of Science and Technology MISIS, Moscow, Russia}
	
	\author{E.V. Il’ichev}
	\affiliation{Leibniz Institute of Photonic Technology, 07745 Jena, Germany}
	\affiliation{Russian Quantum Center, Skolkovo village, Russia}
	
	\author{O.V. Astafiev}
	\affiliation{Skolkovo Institute of Science and Technology, Moscow, Russia}
	\affiliation{Moscow Institute of Physics and Technology, Dolgoprundiy, Russia}
	\affiliation{Physics Department, Royal Holloway, University of London, Egham, Surrey TW20 0EX, United Kingdom}
	\affiliation{National Physical Laboratory, Teddington, TW11 0LW, United Kingdom}
	
	\date{October 2020}

	\begin{abstract}
		
		We present a blueprint for a maser with a single three-level transmon superconducting artificial atom. The system can be pumped coherently via a two-photon process, and to achieve high population inversion,  the relaxation rate of the metastable state is increased via an auxiliary low-Q cavity coupled to a transition between the transmon excited states. We show numerically that such a maser can operate both in the intermediate coupling regime with super-Poissonian photon statistics and in the strong coupling regime, where the statistics is sub-Poissonian. For the former, the maser exhibits thresholdless behavior and for the latter, there is a well-defined pumping threshold. A useful side-effect of the auxiliary resonator is that it allows to overcome the photon blockade effect for the pump, which would otherwise prevent high photon population. Finally, we observe the bistability of the steady-state Wigner function and the self-quenching effect for some parameters. 
		
	\end{abstract}
	
	\maketitle

	\section{Introduction}
	
	Conventional multi-atom lasers operate in the regime of weak coupling between the atoms and the light modes, possess a threshold and emit classical light. Single-atom lasers with strong coupling have been gaining increased attention since 1990s, as their behavior was predicted to  be drastically different from usual lasers \cite{oneatomlasers, Ginzel, Pellizzari, Pellizzari1, Horak, Loffer, Meyer, Meyer1, phstat, comment, Kilin, Nori}.  Among their non-standard features are presence or absence \cite{comment, phstat} of the pumping threshold depending on the parameters \cite{natureblatt}; self-quenching, when the cavity population decreases when the pumping rate is increased \cite{oneatomlasers}; photon blockade prohibiting coherent pumping of more than a single photon in the strong-coupling regime \cite{blockade, natureblatt, blockade97, blockadequbits1, blockadequbits2}; bistability of the Wigner function \cite{wignersset, multistability}; sub-Poissonian statistics of the emitted radiation leading to amplitude squeezing and purely quantum phenomena such as photon antibunching or super-Poissonian statistics with high intensity fluctuations and phase squeezing \cite{oneatomlasers}. Owing to these peculiar properties, a single-atom laser may be used for a wide range of applications as a source of non-classical radiation \cite{NielsenChuang,squeezedlight, qmeasurements, qimaging,qimaging1,qcommunications}. 
	
	Experimentally, the first single-atom maser has been implemented in 1985 on Rydberg atoms \cite{firstmaser}. Then, several other platforms for single-atom lasing have been proposed, including trapped ions \cite{natureblatt, iontrap}, trapped Cs atom \cite{firstCs,Cs}, quantum dots \cite{bistabledots, thresholddyn, Nomura}, superconducting single-electron transistors \cite{wignersset, Astafiev} and flux qubits \cite{Neilinger, fluxqubit, qubitampl}. The single-atom masers based on superconducting cQED devices are particularly interesting since they can be used in tandem with superconducting artificial atoms \cite{guide}.
	
	In this work, we present a blueprint for a maser in this architecture using a transmon-type artificial atom \cite{koch}. The transmon may be regarded as a $\Xi$-system with the eigenstates $\ket{g}, \ket{e} ,\ket{f}$ referred to the ground, first and second excited states, respectively. The lasing transition $\ket{e} \rightarrow \ket{g}$ is in resonance with the reservoir lasing cavity coupled to the transmon. A classical lasing scheme in quantum optics is based on a Lambda-configuration of a three-level atom \cite{oneatomlasers}. However, transmon is a cascade system and cannot be excited directly to the second excited state, and thus the pumping is done via two-photon excitation of the $\ket{g} \leftrightarrow \ket{f}$ transition. The pumping processes for population inversion may be described as $\ket{g} \xleftrightarrow{2\hbar\omega} \ket{f} \rightarrow \ket{e}$. However, due to the small non-linearity of the transmon the $ \ket{f} \rightarrow \ket{e} $ process is only $\sqrt{2}$ times faster than $ \ket{e} \rightarrow \ket{g} $. Therefore, to increase the effective pumping rate, we suggest to couple the transmon to another low-Q cavity resonant with the $\ket{f} \rightarrow \ket{e}$ transition, very similar to what was done before for a chain of transmons \cite{ma2019dissipatively}. This cavity plays a role of artificial bath \cite{cavitybath, artificialbath, bathengineering, bathengineering1}, which relaxes the system to $\ket{e}$, ensuring the metastability of $\ket{f}$. Then, we solve the master equation numerically for the full system and study its transient and steady-state behavior for various combinations of parameters.
 	
 	The suggested model is different from those studied before at least in two ways. First, we use a two-photon process for coherent pumping. Second, we employ engineered $\ket{f} \rightarrow \ket{e}$ dissipation via an auxiliary resonator, which at the same time significantly modifies the system spectrum. Actually, the latter fact alleviates the effect of the photon blockade \cite{blockade} caused by the large size of the vacuum Rabi splittings in the strong coupling regime \cite{Pellizzari}, similar to the case of the cascade laser proposed in 1995 \cite{cascade}. This effect allows to pump more than 40 photons in the strong-coupling regime. A similar breakdown of the photon blockade was investigated before in a simpler Jaynes-Cummings system \cite{BlockadeBreakdown}. 
	
	Additionally, our simulations predict that depending on its parameters, the system can be operated in two distinct regimes. In one regime we observe a well-defined lasing threshold in our architecture which we also associate with the unusual energy level structure and blockade breakdown. In the other, no threshold is present, and there is a phase transition between the regimes with a sharp increase of the steady-state cavity population and the intensity fluctuations. A similar transition between different lasing regimes was previously predicted for the strong coupling \cite{wignersset, phases}.

	\section{Modeling the system}\label{sec:modeling}
	
\begin{figure}[]
		\centering\includegraphics[width=0.5\textwidth]{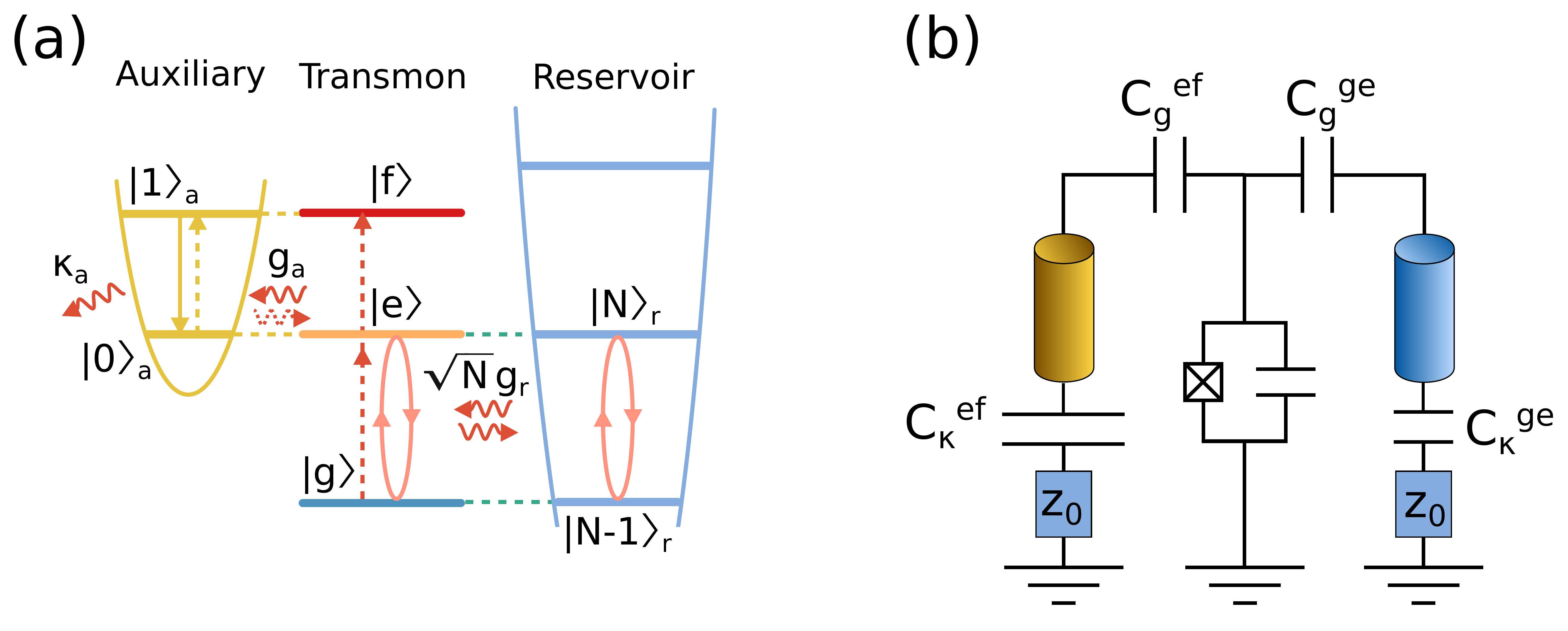}
		\caption{\textbf{(a)} Schematic configuration of the correctly aligned energy levels without coupling between subsystems. The reservoir is resonant with the $\ket{g} \leftrightarrow \ket{e}$ transition, and the auxiliary cavity is resonant with the $\ket{e} \leftrightarrow \ket{f}$ transition. $\ket{g},\ket{e},\ket{f} $ are in blue (medium gray), orange (light gray) and red (dark gray), respectively. \textbf{(b)} Schematic of the circuit. To the left is the auxiliary and to the right is the reservoir cavity. $C_g^{ge}$ ($C_g^{ef}$) is a capacitor that couples transmon with reservoir (auxiliary) cavity, and $C_{\kappa}^{ge}$ ($C_{\kappa}^{ef}$)~-- capacitor that provides the desired external quality factor}\label{scheme}
	\end{figure}

	The scheme of the system is shown conceptually in \autoref{scheme}~(a). A tunable transmon is coupled to two cavities: the reservoir (right) and the auxiliary (left) with the coupling strengths $g_{r}$ and $g_{a}$, respectively. The reservoir has high internal and external Q-factors and should accumulate a considerable number of photons being in resonance with the $\ket{g} \leftrightarrow \ket{e}$ transition of the transmon, at frequency $\omega_{ge}$. The second low-Q cavity is resonant with the $\ket{e} \leftrightarrow \ket{f}$ transition at frequency $\omega_{ef}$, and thus provides an engineered metastability of the $\ket{f}$-state. The cavities do not interact directly with each other. To describe this system we apply the Jaynes-Cummings model \cite{jaynes-cummings} with an obvious modification to include two cavities, and in the rotating wave approximation (RWA) the resulting Hamiltonian reads:
	\begin{align}
	\label{hamilt}
	  \hat{H} &=  \hat{H}_{t} + \hat H_{d} + \sum_{\lambda=r,a}( \hat{H}^{(\lambda)}_{c} +  \hat{H}^{(\lambda)}_{i}),\\
	  \hat{{H}}_{t} &= \hbar \Delta
	  b^{\dagger} b +\frac{\hbar \alpha}{2}
	  b^{\dagger} b( b^{\dagger} b-1),\\
	  \hat{H}_{d} &= \frac{\hbar \Omega}{2} ( b + b^\dag),\\
	  \hat{H}^{(\lambda)}_{c}& = \hbar\Delta_c^{(\lambda)} (a_{\lambda}^\dag a_\lambda + 1/2),\\
	  \hat{H}^{(\lambda)}_{i} &= \hbar g_{\lambda}  ( b a_{\lambda}^\dag + b^\dag  a_\lambda),
	\end{align}
	
	\noindent where $\hat{H}_{t},\ \hat H_{d},\ \hat{H}^{(\lambda)}_{c}$ and $\hat{H}^{(\lambda)}_{i}$ are transmon, drive, cavity and interaction Hamiltonians, respectively, and $a_{r}$, $ a_{a}$, $b$ are the standard bosonic annihilation operators for the reservoir, auxiliary resonator and the transmon, respectively. The transmon parameters are: $\Delta = \omega_{ge} - \omega_d$, the detuning of drive frequency $\omega_d$ from the $\ket{g} \leftrightarrow \ket{e}$ transition; $\alpha < 0$, the anharmonicity; $\Omega$, the microwave drive amplitude. Next, $\Delta_c^{(r, a)} = \omega_{r,a} - \omega_d$ are the detunings of the reservoir and auxiliary cavities, respectively. In the ideal configuration, when the reservoir is resonant with the $\ket{g} \leftrightarrow \ket{e}$ transition, the auxiliary cavity is resonant with the $\ket{e} \leftrightarrow \ket{f}$ transition, and the drive is at the two-photon frequency $\omega_d = \omega_{gf}/2 = \omega_{ge} + \alpha/2$, we can find $\Delta_c^{r} = \Delta = - \alpha/2$ and $\Delta_c^{a} = \alpha/2$.

	Since relaxation processes are essential for lasing, modeling the system requires solution of the full Lindbladian master equation: for the transmon, the reservoir, and the auxiliary cavity, we use the relaxing collapse operators $\sqrt{\gamma} \hat b,\ \sqrt{\kappa_{r}} \hat a_{r},\ \sqrt{\kappa_{a}}\hat a_{a}$, respectively. 
	
	To demonstrate the feasibility of the proposed device, we evaluate also its physical parameters. The optimal values for $g_{r, a},\ \kappa_{r, a},\ \omega_d$ were found with a Nelder-Mead algorithm aimed at maximizing the emitted power from the reservoir calculated by the input-output theory \cite{gardiner1984master}:
	\begin{equation}\label{power}
		P = \kappa_{r} N_{ss} \cdot \hbar \omega_{ge},
	\end{equation}
	where $N_{ss}$ is the steady-state number of photons in the reservoir.
	\begin{figure*}[]
		\centering\includegraphics[width=1\textwidth]{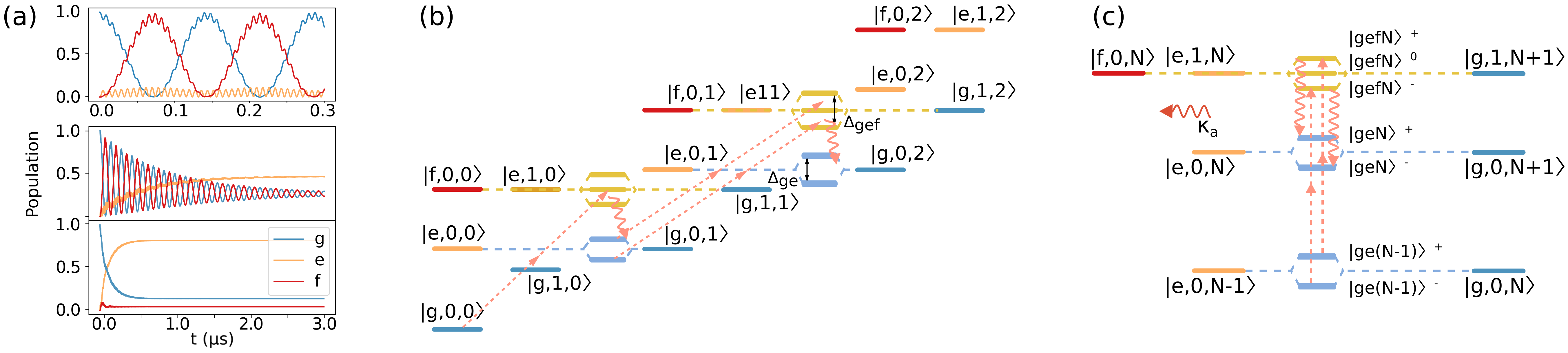}
		\caption{ \textbf{(a)} Population of the transmon levels under a two-photon drive, $\Omega/2\pi = 25$ MHz. Upper panel: $\gamma = 0$, $g_{a} = 0$, middle panel: $\gamma = 1$ $\mu\text{s}^{-1}$, $g_{a} = 0$, lower panel: $\gamma = 1$ $\mu\text{s}^{-1}$, $g_{a}/2\pi = 8$ MHz, $\kappa_{a} = 138$ $\mu\text{s}^{-1}$.
		\textbf{(b)} Low-energy subspace of the system spectrum. Dashed arrows show two-photon pumping, wavy arrows~-- relaxation processes. Blue (medium gray) levels correspond to $\ket{g}$, orange (light gray)~-- $\ket{e}$, red (dark gray)~-- $\ket{f}$.
		\textbf{(c)} A single pumping stage showing transition from $N-1$ to $N$ photons in the reservoir and the participating eigenstates.}\label{sim}
	\end{figure*}
	In our simulations, truncating the reservoir Fock space at 60 photons, we find the optimal parameters to be the following: $g_{r}/2\pi = 6.5$ MHz, $g_{a}/2\pi = 23.5$ MHz, $\kappa_{r} = 0.31$ $\mu\text{s}^{-1}$, $\kappa_{a} = 138$ $\mu\text{s}^{-1}$ and $\omega_d = \omega_{gf}/2$, for $\Omega/2\pi = 25$ MHz. The algorithm converges to the same values starting from different initial parameters, so there is probably a single optimum in the parameter space. Since $\kappa_{a}$ is much larger than the two-photon Rabi frequency (see below for details), the number of photons in auxiliary cavity stays close to zero on average, so we truncate its Fock space to one photon. For the optimization, we use fixed realistic values for $\gamma = 0.1$ $\mu\text{s}^{-1}$, $\omega_{ge}/2\pi = 6$ GHz, $\alpha/2\pi = -200$ MHz, $\omega_r/2\pi = 6$ GHz, and $\omega_a/2\pi = 5.8$ GHz. For the optimal parameters we found $N_{ss} = 49$, which corresponds to $-132$ dBm of emitted power.  
	
	Fabricating the superconducting cavities and the transmon with megahertz accuracy in frequency and anharmonicity should be experimentally attainable, and then the ideal configuration can be reached by tuning the transmon into the resonance with the reservoir via the external magnetic flux. Schematically, the circuit model of the device is shown in \autoref{scheme}~(b).
	
	The reservoir population is tolerant to small changes of the parameters, which can occur during fabrication. $N_{ss}$ becomes lower, but still enough to measure, due to mismatch between the auxiliary resonator frequency and $\ket{e} \leftrightarrow \ket{f}$ transition up to 30 MHz, while $\ket{g} \leftrightarrow \ket{e}$ frequency of the transmon can be always tuned directly to resonance with the reservoir cavity. The changes of $\kappa_{a}$ do not affect $N_{ss}$ significantly: higher $\kappa_{a}$ does not change $N_{ss}$ at all, while up to 40\% lower $\kappa_{a}$ still allows for pumping enough photons to measure. Variation of $g_{r}$ and $g_{a}$ affects the population significantly but slowly (see below for details), so their deviation from ideal values also should not be a problem. The only parameter whose deviation significantly affects the maser is $\kappa_{r}$: the lower $\kappa_{r}$ allows for pumping much more photons, but higher $\kappa_{r}$ may completely prohibit lasing. However, higher $\kappa_{r}$ can be coped with by increasing $g_{r}$.
	
	In addition, we have verified that adding dephasing to the transmon with $\gamma_\phi$ in the range of one $\mu\text{s}^{-1}$ and even tens of $\mu\text{s}^{-1}$ does not alter the number of pumped photons significantly.

	\section{Pumping dynamics}\label{sec:splittings}
	
	To calculate numerically the dynamics of the system, we use the QuTiP Python package \cite{qutip}. Below we denote the factorized states of the tripartite system as $\ket{\mu, M, N}$ where $\mu$ is the transmon state ($g$, $e$, $f$), and $M,\ N$ are the number of photons in the auxiliary resonator and the reservoir, respectively. To facilitate the interpretation of the simulation results, we split the task into several steps. 
	
	First of all, we simulate the transmon under a two-photon drive by turning off the coupling to the cavities and the dissipation. The simulated evolution is shown in the upper panel of \autoref{sim}~(a) (here and below we truncate the transmon subspace to four states). One can clearly see two-photon Rabi oscillations when the $\ket{f}$ energy level is fully populated by the two-photon process. There are also small oscillations of the $\ket{e}$-level population from the virtual processes.
	
	In the middle panel \autoref{sim}~(a), we add natural relaxation to the transmon. The rate of the $\ket{f} \rightarrow \ket{e}$ transition is higher $n_{12}/n_{01}$ times than of the $\ket{e} \rightarrow \ket{g}$ one, where $n_{ij}$ is matrix element of charge operator in the basis of transmon eigenstates \cite{koch}. In this case, some population inversion may be achieved without any additional effort. Nevertheless, when the coupling to the auxiliary cavity it turned on, the population inversion becomes significantly larger, as shown in the lower panel of \autoref{sim}~(a).
	
	Finally, we turn on the coupling to the reservoir cavity. The resulting level structure of the full tripartite system is depicted in \autoref{sim}(b), and \autoref{sim}(c) shows the core process of pumping an additional photon in the reservoir. One can see that there are two types of Rabi splittings in the spectrum. The first, caused by the interaction between the reservoir and the $\ket{g} \leftrightarrow \ket{e}$ transition of the transmon, leads to the formation of the 
	\begin{equation}
		\ket{geN}^\pm = \frac{1}{\sqrt{2}}(\ket{g,0,N+1} \pm \ket{e,0,N})
	\end{equation}
	dressed states composed of the factorized states with different reservoir populations. To avoid ambiguity, we label them using the lower number $N$. The size of this splitting is calculated as
	\begin{equation}\label{splitting1}
		\Delta_{ge}(N) = 2 g_{r} \sqrt{N+1}.
	\end{equation}
	
	The second splitting is due to the coupling between three degenerate states $\ket{f,0,N}$, $\ket{e,1,N}$, and $\ket{g,1,N+1}$. When the degeneracy is lifted, triplet eigenstates appear:
	\begin{equation}
	\begin{gathered}
		\ket{gefN}^0 = \sin\theta \ket{g,1,N+1} - \cos\theta \ket{e,1,N},
	\end{gathered}
	\end{equation}
	\begin{equation}
	\begin{gathered}
		\ket{gefN}^\pm = \frac{1}{\sqrt{2}} (\cos\theta \ket{g,1,N+1} \pm \ket{e,1,N} + \\ + \sin\theta \ket{f,0,N}),
	\end{gathered}
	\end{equation}
	\noindent where
	\begin{equation}
		\tan\theta = \frac{g_{a}}{g_{r}\sqrt{N+1}}
	\end{equation}
	is the mixing angle. The energy difference between $\ket{gefN}^+$ and $\ket{gefN}^-$ is
	\begin{equation}\label{splitting2}
		\Delta_{gef}(N) = 2 \sqrt{(N+1) g_{r}^2 + g_{a}^2}.
	\end{equation}
	\noindent As before, we label the triplet states by the lowest $N$ among their factorized components. 
	
	We note that the first splitting (\autoref{splitting1}) is ordinary for any coupled atom-cavity system and grows as $\sqrt{N}$ with photon population of the cavity \cite{sqrtN}. In contrast, the second one (\autoref{splitting2}) is the characteristic feature of our system due to the presence of the auxiliary resonator and a certain frequency configuration. As we show below, both these splittings play the crucial role in the dynamics and allow the device to function in principle for arbitrarily large reservoir populations.

	\section{Simplified analytical solution}\label{sec:approx}
	
	\begin{figure}[]
		\centering \includegraphics[width=0.5\textwidth]{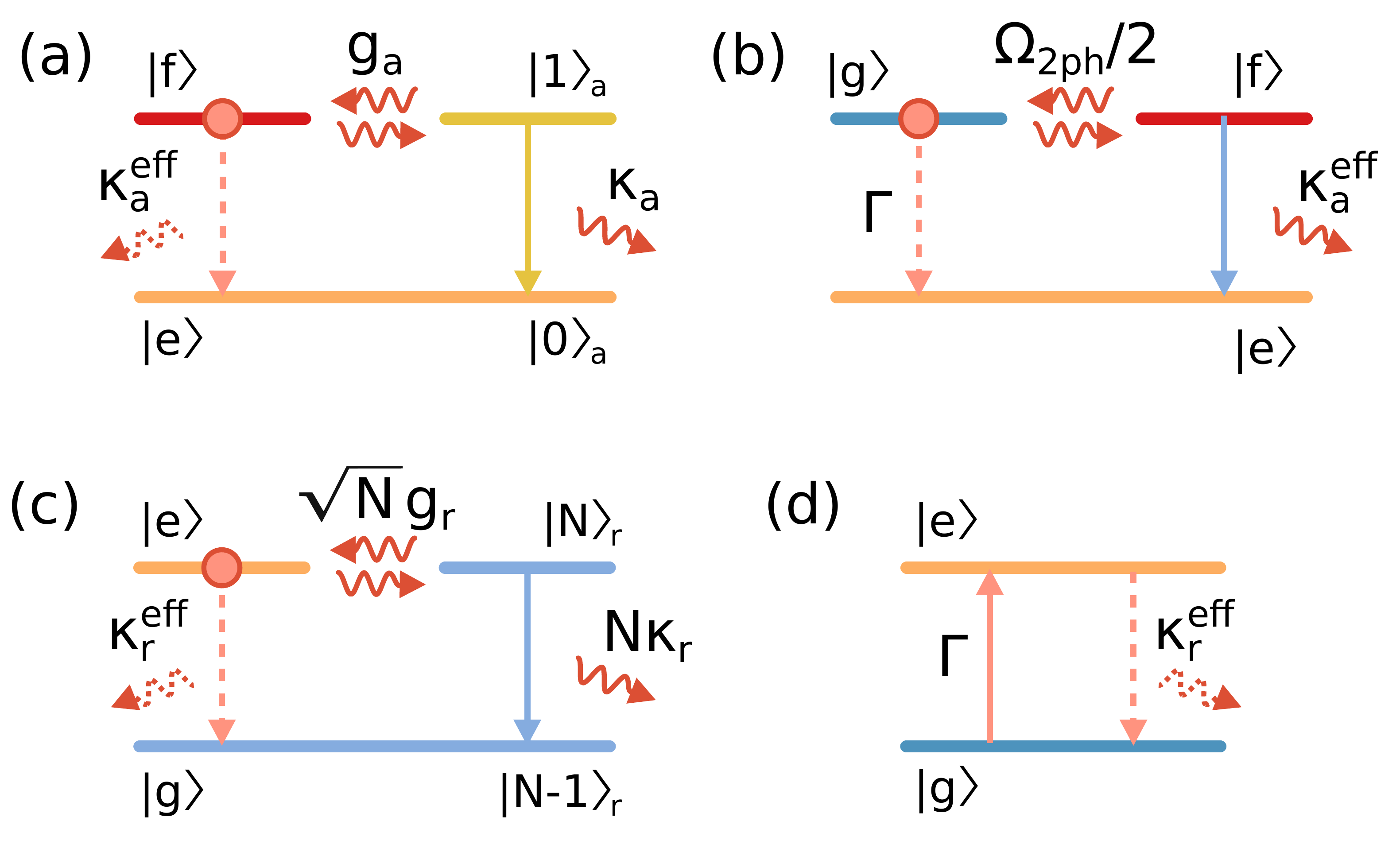}
		\caption{\textbf{(a)} Estimating the effective relaxation rate $\kappa_a^\text{eff}$ due to the auxiliary cavity. \textbf{(b)} Estimating the effective pumping rate due to the two-photon drive $\Omega_{2ph}$ and the effective relaxation $\kappa_a^\text{eff}$. \textbf{(c)} Estimating the effective relaxation rate due to the reservoir with $N$ photons inside. \textbf{(d)} $\ket{g}$ and $\ket{e}$ levels of transmon with effective $\ket{e} \rightarrow \ket{g}$ dissipation, which act as a two-level laser.}\label{analytical}
	\end{figure}
	
	To calculate approximately the reservoir population in the steady state, we consider only the $\ket{g}$ and $\ket{e}$ states with the effective pumping rate $\Gamma$ and effective dissipation rate $\kappa_r^\text{eff}$ through the reservoir (\autoref{analytical}(d)). In this simplified view, the balance condition $\kappa_r^\text{eff} = \Gamma$ allows to determine the reservoir population when the pumping ceases since $\kappa_r^\text{eff}$ is proportional to $N$ and $\Gamma$ is constant. We prove this in several steps below.
	
	Firstly, we calculate the effective $\ket{f} \rightarrow \ket{e}$ relaxation rate due to the auxiliary cavity (\autoref{analytical}(a)). Consider two coupled two-level systems: the first two-level system (on the left) represents $\ket{e}$ and $\ket{f}$ levels of transmon, and the second one (on the right)~-- $\ket{0}$ and $\ket{1}$ levels of the auxiliary resonator. The relevant master equation for a Hamiltonian $H = g_{a} (b a_{a}^\dag + a_{a} b^\dag)$ with a collapse operator $\sqrt{\kappa_a} a_{a}$ assuming initial state $\ket{f}$ has separate analytical solutions for the underdamped and overdamped regimes. 
	
	In the underdamped, or strong coupling regime when $4 g_{a} > \kappa_a$, the population of $\ket{f}$ is
	\begin{equation}
	\rho_{ff} = \frac{e^{-\frac{\kappa_a t}{2}}}{\cos^2\theta} \cos^2(g't+\theta),
	\end{equation}
	\noindent where $g' = \sqrt{16 g_{a}^2 - \kappa_a^2}/4$, $\tan \theta = \frac{\kappa_a}{g'}$. Using the decay rate of the vaccuum Rabi oscillations, one may assume that
	\begin{equation}\label{kappa_strong}
	\kappa_a^\text{eff} = \frac{\kappa_a}{2}.
	\end{equation}
	
	In the overdamped, or weak coupling regime where $4 g_{a} < \kappa_a$,
	\begin{equation}
	\rho_{ff} = \frac{e^{-\frac{\kappa_a t}{2}}}{\cosh^2\theta} \cosh^2(\gamma' t+\theta),
	\end{equation}
	\noindent where $\gamma' = \sqrt{\kappa_a^2 - 16 g_{a}^2}/4$, $\tan \theta = \frac{\kappa_a}{\gamma'}$.
	The decay rate is determined by the slowest exponent, so
	\begin{equation}\label{kappa_weak}
	\kappa_a^\text{eff} = \frac{\kappa_a - \sqrt{\kappa_a^2 - 16 g_{a}^2}}{2}.
	\end{equation}
	
	One can see that while $g_{a}$ is fixed, increasing $\kappa_a$ at first increases but then decreases $\kappa_a^\text{eff}$ upon transition from the underdamped to the overdamped regime ($\kappa_a^\text{eff} \rightarrow 4 g_a^2/\kappa_a$ there). Physically, this effect can be explained by the broadening of the $\ket{1}_{a}$ level, which prevents transfer of energy from $\ket{f}$.
	Thus, $\kappa_a^\text{eff}$ reaches its maximum when $\kappa_a = g_{a}$ which is consistent with our numerical simulations.
	
	Next, we calculate effective $\ket{g} \rightarrow \ket{e}$ pumping rate $\Gamma$. The simplified model of the system with two-photon pumping and $\ket{f} \rightarrow \ket{e}$ dissipation is shown on \autoref{analytical}(b). The two-photon process is replaced by the equivalent one-photon drive of strength $\Omega_{2ph}$ (see Appendix \ref{rate} for details). This step is only approximate as it does not take into account the full complexity of the energy spectrum and the possibility of the off-resonant two-photon drive. 
	From \autoref{analytical}(b) one can see that the tree-level system with pumping and dissipation is completely equivalent to the two coupled two-level systems in \autoref{analytical}(a) with $g_{a}$ replaced by $\Omega_{2ph}/2$. The solution for the such a system is the same: in the strong coupling regime $2 \Omega_{2ph} > \kappa_a^\text{eff}$ the effective pumping rate is
	\begin{equation}
		\Gamma = \frac{\kappa_a^\text{eff}}{2},
	\end{equation}
	\noindent and for the weak coupling regime $2\Omega_{2ph} < \kappa_a^\text{eff}$
	\begin{equation}\label{gamma_eff}
	\Gamma = \frac{\kappa_a^\text{eff} - \sqrt{(\kappa_a^\text{eff})^2 - 4 \Omega_{2ph}^2}}{2}.
	\end{equation}
	
	In our case, the strong coupling is not achievable due to the weakness of the two-photon drive. Consequently, we take \autoref{gamma_eff} to calculate effective pumping rate. For the optimal parameters from \autoref{sec:modeling}, it leads to $\Gamma = 5.3$ MHz.
	
	The next step is the calculation of effective dissipation rate $\kappa_r^\text{eff}$ due to the reservoir populated with $N$ photons (\autoref{analytical}(c)). Again, the system is equivalent to \autoref{analytical}(a). Coupling strength $g_{r}$ should not be much lower then $g_{a}$, otherwise the Rabi splittings will disturb pumping (see next section for detail). Assuming this, for optimal $g_{a}$ and $\kappa_r<1$ $\mu \text{s}^{-1}$, the condition for the strong coupling regime $4\sqrt{N}g_{r} > N\kappa_r$ is met even for high reservoir populations ($N \sim 100$). Therefore,
	\begin{equation}
		\kappa_r^\text{eff} = \frac{N_{ss} \kappa_r}{2}.
	\end{equation}
	
	The energy balance condition is $\kappa_r^\text{eff} = \Gamma$ (\autoref{analytical}(d)). Consequently, the maximum number of pumped photons can be expressed as 
	\begin{equation}
		N_{ss} = 2\frac{\Gamma}{\kappa_r}.
	\end{equation}
	
	For the optimal parameters, this equation gives $N_{ss} = 34$. The exact simulated value is $N_{ss} = 49$. The 30\% difference is probably caused by the differences in the real level structure due to the coupling to both cavities.

	\section{Overcoming photon blockade}\label{sec:blockade}
	
	The simplified model studied in the previous section does not take into account the effects of the photon blockade as it replaces the coherent pumping by an effective incoherent one. However, the auxiliary cavity actually allows to treat the system within the simplified model. To see this, one should first consider a simple device that does not include an auxiliary resonator, operates in the strong-coupling regime and is subject to photon blockade as follows. Due to the $\sqrt{N}$ growth of the $\Delta_{ge}(N)$, there is no single frequency that would satisfy the resonance condition of the two-photon pumping for an arbitrary population of the reservoir. The pumping at the $\ket{g,0,N} \leftrightarrow \ket{f,0,N}$ frequency becomes off-resonant when $\langle \hat N \rangle = \langle a_{r}^\dag a_{r} \rangle \geq 1$ \cite{Pellizzari} (see \autoref{sim}~(b)) as there would be no triplet Rabi splittings and the detuning would be $\delta = \Delta_{ge}(N)/2$. This effect is observed for all coherently pumped single-atom lasers: $\delta = g_{r} \sqrt{N} \rightarrow \infty$ for $N \rightarrow \infty$ \cite{blockade}.
	
	However, in the presence of the auxiliary resonator one can see from \autoref{sim}~(b,c) that the detuning of the two-photon drive is instead calculated as
	
	\begin{equation}\label{detuning}
	\delta = \left[\Delta_{gef}(N+1)-\Delta_{ge}(N)\right]/2
	\end{equation}
	
	One can see that $\delta \rightarrow 0$ for $N \rightarrow \infty$ (\autoref{sim}(c)), which means that it would be possible to use monochromatic pumping at $\omega_{gf}/2$ for sufficiently large values of $\langle \hat N \rangle$, even though it would be slightly off-resonant for low $\langle \hat N \rangle$. In other words, if it is possible to pump several photons in the reservoir off-resonantly, then the subsequent pumping will become resonant and similar to the simplified model. We demonstrate numerically that the latter is possible due to the large dissipative bandwidths of the states in the $\ket{gefN}^{\pm,0}$ splittings. The low Q-factor of the auxiliary resonator allows off-resonant transitions to them even if the detuning is of the order of several MHz. In \autoref{pumping}~(a, b), we compare the steady-state solution with the transient solutions at 1 an 2 $ \mu s $. The steady-state picture shows that pumping at a frequency close to $\ket{g,0,0} \leftrightarrow \ket{f,0,0}$ is indeed optimal in the long term, despite that at first the sideband $\ket{geN}^+ \leftrightarrow \ket{gefN}^+$ has a higher pumping rate, see Appendix \ref{rate} for details; the sidebands are subject to the photon blockade and eventually die off for large $\langle \hat N \rangle$.
	
	As a result, with optimal parameters, we can pump $\approx 50$ photons while maximizing output power (-132 dBm at the optimum). Maximizing the reservoir population allows pumping at least 90 photons in the steady state (above this point, the simulation becomes intractable due to the size of the necessary Hilbert space).

	\section{Different lasing regimes and phase transitions}\label{sec:regimes}
	
	While the simplified model is applicable for large reservoir populations, numerical solution is required to accurately characterize the behavior of the system for arbitrary parameter combinations. We find that the most important parameters are the coupling strengths between the transmon and the cavities, so in this section we study in detail how they affect the reservoir steady state properties.
	
	\begin{figure}[]
		\centering \includegraphics[width=0.5\textwidth]{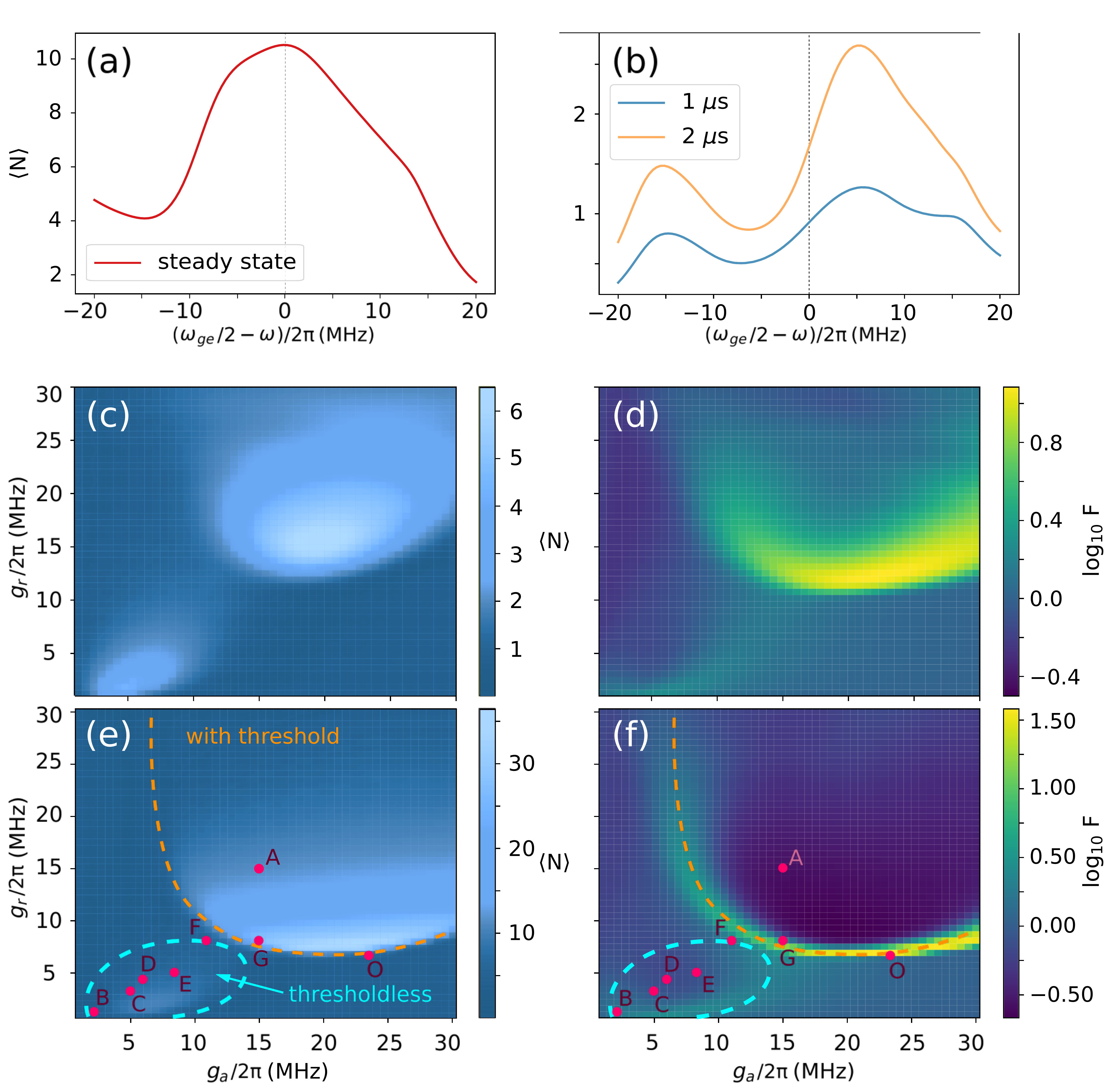}
		\caption{\textbf{(a, b)} Average number of photons in the reservoir depending on the drive frequency $\omega$ for $g_{r}/2\pi = g_{a}/2\pi = 15$ MHz, $\kappa_{r} = 0.2$ $\mu\text{s}^{-1}$, $\kappa_{a} = 128$ $\mu\text{s}^{-1}$, $\Omega/2\pi = 20$ MHz (point A in (e, f)) in the steady state (a) and in 1 $\mu$s and 2 $\mu$s after the pumping was started (b). \textbf{(c, e)} Pumped photons and \textbf{(d, f)} logarithm of Fano factor vs. coupling strengths for $\kappa_{a} = 138$ $\mu\text{s}^{-1}$, $\kappa_{r} = 0.2$ $\mu\text{s}^{-1}$. \textbf{(c, d)} $\Omega/2\pi = 15$ MHz, \textbf{(e, f)} $\Omega/2\pi = 20$ MHz. The letters are different sets of the parameters used in the other subsections (O depicts the optimal parameters). Orange dashed line marks the transition to the threshold-bearing regime and blue dashed line the thresholdless regime.}\label{pumping}
	\end{figure}
	
	\autoref{pumping}(c-f) shows the reservoir population and the logarithm of the Fano factor $F = \frac{\langle N^2 \rangle - \langle N \rangle^2}{\langle N \rangle}$ in the steady state depending on $ g_{r, a} $ for two different values of $\Omega$. $F$ is a measure of intensity fluctuations of the laser field: $F<1$ indicates sub-Poissonian statistics and photon antibunching, $F=1$~-- coherent Poissonian field, and $F>1$~-- super-Poissonian statistics with high intensity fluctuations.
	
	For both driving strengths, one can see an area with a significant increase in the $N_{ss}$ and $F$ when the coupling strengths $ g_{r,a} $ become large enough. We find that this area marks the phase transition between two regimes of lasing: threshold-bearing and threshold-less (see \autoref{threshold-sec} for detail). The lasing threshold is said to exist if there is some pumping power value for which a rapid increase of intensity fluctuations and usually of the steady-state photon number in the cavity is observed \cite{comment}.
	
	\begin{figure*}[]
		\centering\includegraphics[width=1\textwidth]{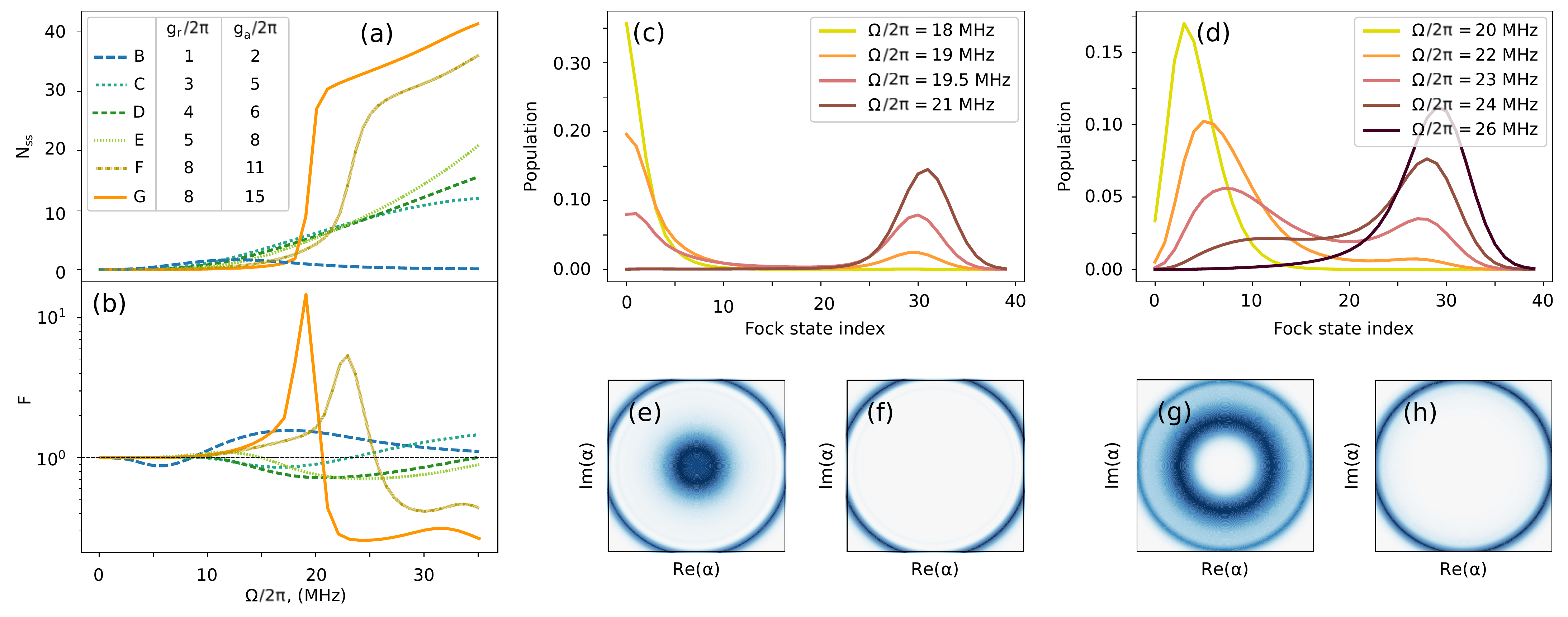}
		\caption{ Reservoir population \textbf{(a)} and the Fano-factor \textbf{(b)} for the threshold-less and threshold-bearing regimes (points B-F in \autoref{pumping}, values for $g_{a,r}/2\pi$ in the legend are in MHz). For all panels $\kappa_{a} = 138$ $\mu\text{s}^{-1}$, $\kappa_{r} = 0.2$ $\mu\text{s}^{-1}$. \textbf{(c-d)} Reservoir Fock state probability for various $\Omega$: \textbf{(c)} point G, \textbf{(d)} point F. \textbf{(e-h)} Steady-state reservoir Wigner functions: \textbf{(e)} for (c), $\Omega/2\pi = 19.5$ MHz;  \textbf{(f)} for (c), $\Omega/2\pi = 21$ MHz;
	    \textbf{(g)} for (d), $\Omega/2\pi = 23$ MHz;
	    \textbf{(h)} for (d), $\Omega/2\pi = 26$ MHz.}\label{threshold}
	\end{figure*}
	
	For the smaller drive amplitude, the phase transition is not very pronounced; however, for the larger drive, it is sharp (orange dashed line). After the transition, the lasing is characterized by the large $ N_{ss} $ and $F<1$. Finally, the area of the threshold-bearing regime grows when the pumping rate is increased.
	
	Qualitatively, the location of the threshold-bearing regime in the parameter space can be explained as follows. Firstly, the large coupling constant $g_{r}$ allows pumping the reservoir quicker, which explains why the transition requires $g_{r}/2\pi$  to exceed 12 and 8 MHz, for the weak and strong drive, respectively. Secondly, when $g_{a}/2\pi < 8$ MHz, the coupling to the auxiliary resonator is too low to reach significant $\kappa_{a}^\text{eff}$ and to produce sufficient population inversion. Finally, increasing $g_{a}$ while  keeping $g_{r}$ fixed reduces $N_{ss}$ due to the increasing detuning discussed in \autoref{sec:blockade}.
	
	In the thresholdless regime in the bottom left corner of \autoref{pumping}(c, e) (dashed blue line)  it is still possible to pump a considerable number of photons (about 5-10) while $F \gtrsim 1$ in this area. This regime is standard for single-atom lasers in the intermediate and low-coupling regime. The coupling constants in the thresholdless regime are low, so the energy splittings $\Delta_{ge}(N)$ and $\Delta_{gef}(N)$ do not play any role due to the dissipative line broadening.

	\section{Threshold and bistability}\label{threshold-sec}
	
	Single-atom lasers previously discussed in literature were thresholdless in the strong coupling regime \cite{natureblatt, firstCs,comment}, but could have one or two thresholds in the regime of weak and intermediate coupling \cite{natureblatt, iontrap, coher}. Therefore, below we discuss why in our case it is possible to observe a threshold even in the strong coupling regime.
	
	\autoref{threshold}(a, b) shows the steady-state values for $N_{ss}$ and $F$ for five different combinations of coupling constants, corresponding to points B-F in \autoref{pumping} (e, f), depending on the drive amplitude $ \Omega $. For the first four sets of parameters (B-E), there is no well-defined threshold, even for $g_{r}/2\pi = 5$ MHz, $g_{a}/2\pi = 8$ MHz (E), which has a small maximum of $F$ at $\Omega/2\pi = 12$ MHz. The photon distribution in the regimes B-E can be super-Poissonian as well as sub-Poissonian depending on the drive power. The observed behavior is consistent with previous works \cite{phstat, coher}.
	
	For the combination B, $g_{r}/2\pi = 1$ MHz and $g_{a}/2\pi = 2$ MHz, we observe self-quenching probably caused by the dressing of the energy levels by the strong drive. This effect is still not clear to us, because self-quenching remains even if we manually tune pumping frequency into resonance with the correct transition to one of the dressed states, even though the maximum of pumped photons shifts to the larger $\Omega$.
	
	The most interesting feature in \autoref{threshold} (a),(b) is the presence of a well-defined threshold for the combination G: $g_{r}/2\pi = 8$ MHz and $g_{a}/2\pi = 15$ MHz and F: $g_{r}/2\pi = 8$ MHz and $g_{a}/2\pi = 11$ MHz. Our simulations show that such a threshold is present for any point above the orange dashed line at \autoref{pumping} (e, f). Above the threshold, for some parameters there is also no manifestations of the photon blockade~-- the average number of photons increases steadily with growing $\Omega$. As mentioned above, this is different from the previous results for single-cavity systems with coherent pumping in the strong-coupling regime \cite{iontrap,Cs}.
	
	\autoref{threshold}~(c, e, f) illustrate the cavity state in the vicinity of the threshold for the point G, and \autoref{threshold}~(d,g,h)~-- for the point F. One can see bistability which manifests itself as a double-ring structure of the Wigner function. At a sufficiently high driving power for the bistability to vanish, only the outer ring of Wigner function remains and the Fano factor is significantly lower than one.
	
	We explain the bistability in the studied system as follows. We consider the intermediate cavity states with $N$ approximately between 2 and 25 for \autoref{threshold} (c) and between 5 and 25 for \autoref{threshold} (d). These states can not be populated via the coherent pump due to the large detuning they have due to vacuum Rabi splittings (see Eq. \ref{detuning}), and the relaxation dominates their population dynamics. However, the upper states can be pumped resonantly, as discussed before, so the system can remain there indefinitely. Therefore, with increasing drive strength the higher levels gradually absorb more and more of the population from the lower ones as the population is leaking better and better through the ``blocked'' states. In the middle of the threshold transition, this results in the apparent bimodal distribution of photons and very high intensity fluctuations. Additionally, in Appendix \ref{wignerappendix}, we show that bistability would not be observed if the Rabi-splittings did not depend on the photon number. 
	
	This mechanism resembles the one studied in \cite{bistablecoupling}, where the high-order multiphoton processes are becoming closer and closer to each other in frequency, while the single-photon processes are prone to photon blockade. Additionally, the similar mechanism was experimentally confirmed in \cite{BlockadeBreakdown}, where the photon blockade in the strong-coupled qubit-resonator system was broken by the weakly-coupled third level of the qubit, making resonant pumping of the cavity possible.

	\section{Conclusion}
	
   	In this work, we studied theoretically a single-atom maser based on two-photon coherent pumping of a transmon. We find that to attain high population inversion, one needs to artificially increase the relaxation rate of its $\ket{e} \rightarrow \ket{f}$ transition and show that this is feasible by coupling it to an auxiliary low-Q resonator.

    Our master equation simulations predict two distinct regimes of operation for the device with the presence or absence of the lasing threshold. The boundary between the regimes is determined by the coupling strengths between the transmon and the cavities.

    In the threshold-less regime, our system demonstrates non-classical behavior similar to the previously known for single-atom lasers, including both sub-Poissonian and super-Poissonian statistic, and self-quenching.
    
	\begin{figure*}[]
		\centering \includegraphics[width=1\textwidth]{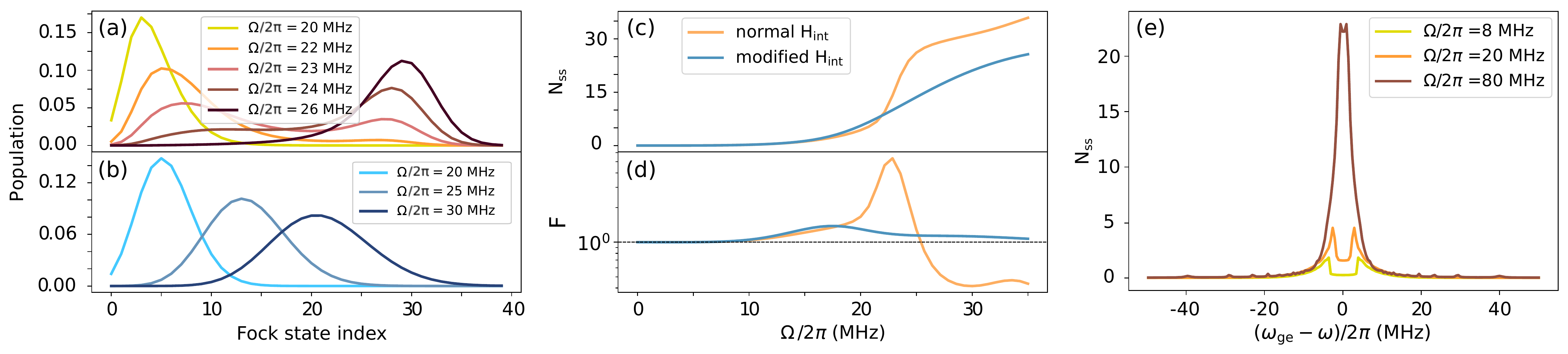}
		\caption{Visualized reservoir states for the system with the following parameters: $\kappa_{a} = 138$ $\mu\text{s}^{-1}$, $\kappa_{r} = 0.2$ $\mu\text{s}^{-1}$, $g_{r}/2\pi = 8$ MHz, $g_{a} = 11$ MHz (point F at \autoref{pumping}(e, f), for different $\Omega$.  \textbf{(a)} With ordinary interaction Hamiltonian, \textbf{(b)} with modified interaction Hamiltonian. \textbf{(c,d)} Number of photons in the reservoir (c) and Fano factor (d) for ordinary and modified interaction Hamiltonian. \textbf{(e)} Number of photons in the reservoir depending on the detuning of the drive frequency $\omega$ from a frequency of reservoir resonator $\omega_{ge}$. The system is modified such a way that allows only one-photon direct reservoir pumping: the number of the transmon states is truncated to 2, when $g_r/2\pi = 8$ MHz, $g_a = 0$ and the other parameters are the same as in \autoref{threshold}. }\label{wignerpeaks}
	\end{figure*}
	
	In the other regime, for which the coupling strengths are higher, we observe a well-defined lasing threshold, marked by an increase of intensity fluctuations. We associate its presence with the interplay between the photon blockade and the peculiarities of the energy spectrum caused by the auxiliary resonator, and the consequent bistability of Wigner function. The light statistics in the reservoir in this regime is highly sub-Poissonian, with $F<0.5$, and its population is of order of tens of photons.
	
	Finally, via numerical optimization, we find an optimal set of parameters for a device that allows pumping about 50 photons in the strong-coupling regime, which corresponds to $-132$ dBm of emitted power. The statistics in this regime is predicted to be sub-Poissonian with $F = 0.27$. We believe that due to the non-classical properties such maser may have a wide range of applications for microwave quantum optics on a chip \cite{qimaging,NielsenChuang,squeezedlight}.

	\section{Acknowledgments}
	
	The investigation was conducted with the support of Russian Science Foundation, Grant No. 16-12-00095. We also acknowledge support from the Ministry of Education and Science of the Russian Federation in the framework of the Increased Competitiveness Program of the National University of Science and Technology MISIS (Contract No. K2-2020-022).

	\appendix
	
	\section{Estimated pumping rate of the two-photon process}\label{rate}
	
	The frequency of the two-photon Rabi oscillations for a three-level system is generally calculated as \cite{twophoton}
	\begin{equation}
	\Omega_{2ph} = \frac{\Omega^2}{2 \Delta},
	\end{equation}
	
	\noindent where $\Delta$ is the detuning between the pump frequency and the intermediate level. In our case, this is the detuning between $\omega_d$ and the splitted energy levels $\ket{geN}^\pm$ (\autoref{sim}(d)). In case of $\ket{g,0,0} \leftrightarrow \ket{f,0,0}$ pumping $\Delta = \alpha/2$. For $\ket{ge0}^{\pm} \leftrightarrow \ket{gef1}^{\pm}$-pumping $$\Delta^{\pm} = \frac{\alpha}{2} \mp \frac{\Delta_{gef}(1)+\Delta_{ge}(0)}{4},$$ which we will call the ``plus" and ``minus" transitions. This leads to the fact that for the parameters in \autoref{pumping}(a, b) the two-photon pumping rate of $\ket{gef0}^+$ level is 1.9 times higher then of $\ket{gef0}^-$ level. Since the pumping rate for the "plus" transition is higher then for the "minus" one for any $N$, in \autoref{pumping}(a, b) the line corresponding to the ``plus'' transition is much brighter compared to the ``minus'' at the beginning of pumping. When $N\rightarrow \infty$, the frequencies of both types of transitions becomes equal, since $\delta \rightarrow 0$ (\autoref{detuning}). So, it is not possible to distinguish them in the steady state picture in \autoref{pumping}(a, b).

	\section{Role of level splittings in bistability}\label{wignerappendix}
	
	In \autoref{wignerpeaks}~(a), the population of the resonator-reservoir for various $\Omega$ is shown. One can see that the intermediate states 5-25 are never populated. We have checked that they are also never populated during the full time evolution while the second peak just grows steadily similarly to what is shown in \autoref{wignerpeaks}~(a) for increasing $\Omega$.
	
	In the main text, we explain the bistability by the large detuning of the intermediate pumping transition. To check if the unusual steady state is really caused by the detuning of these intermediate transitions, we have constructed an artificial interaction Hamiltonian of the reservoir and the transmon. In the new Hamiltonian, the annihilation operator is replaced by a  lowering operator with unity matrix elements so the detuning does not depend on the number of photons in resonator any more. The steady states for such a system are shown in \autoref{wignerpeaks} (b). One can see that now there is only one peak and no bistability.
	
	We have also simulated $ \langle N \rangle $ and $ F $ in the steady state depending on $\Omega$ in \autoref{wignerpeaks} (c,d). One can see that a peak of $F$, which follows the threshold and present for the system with normal $H_{int}^r$ ($F \approx 5$), is absent for the system with modified $H_{int}^r$. Consequently, the threshold, discussed in the main text, is indeed connected with special energy level structure and bistability.

	\section{Direct pumping of the cavities}
	
	One can expect that for large enough $\Omega$ the reservoir can be pumped by direct off-resonant drive, causing $\ket{g,0,N} \rightarrow \ket{g,0,N+1}$ transition. Thus, there could be a restriction on the value of $\Omega$, after which the lasing breaks down. We now check if this supposition is correct.
	
	We set $g_a$ to zero and truncate the number of transmon states to 2. As a result, we have a two-level qubit coupled to the reservoir. Then we drive transmon at the frequency close to $\ket{g}\leftrightarrow\ket{e}$ transition, which is equal to the frequency of the reservoir, and observe the number of photons in the steady state of the reservoir depending on the detuning from this frequency, see \autoref{wignerpeaks}(e). Even for a very strong drive of 80 MHz, the pumping becomes negligible when the detuning is just 20 MHz. In the suggested device, the detuning is $\omega_{ge} - \omega_d = 100$ MHz, so the direct pumping of the reservoir should not be possible even for high $\Omega$.
	
	The direct pumping of the auxiliary cavity is not expected to be possible due to the same reasons, especially given that its relaxation rate is significantly higher than that of the reservoir. We have checked that the line broadening due to the large $\kappa_a$ is not enough to make possible pumping with 100 MHz detuning.
	
	\bibliography{fileapsnew}
	
\end{document}